\documentclass[prx, onecolumn, superscriptaddress, showpacs, nobibnotes, floatfix,amsmath,amssymb,notitlepage,onlinecite]{revtex4-1}

\usepackage{graphicx}
\usepackage{dsfont}
\usepackage{epstopdf}
\usepackage{textcomp}
\usepackage[usenames, dvipsnames]{color}

\graphicspath{ {./figures/} }

\newcommand{\bra}[1]{\ensuremath{\left\langle #1\right|}}
\newcommand{\ket}[1]{\ensuremath{\left| #1\right\rangle}}

\newcommand{\proj}[1]{\ensuremath{\ket{#1}\bra{#1}}}
\newcommand{\mean}[1]{\ensuremath{\left\langle #1\r\rangle}}

\def\us {$\mu$s}

\def\nbare {$\bar{n}_e$}

\def\usinv {$\mu \mathrm{s}^{-1}$}
\def\usinvsp {$\mu \mathrm{s}^{-1}$ }

\renewcommand{\l}[0]{\left}
\renewcommand{\r}[0]{\right}
\newcommand{\cc}{^{\ast}}                						
\newcommand{\hc}{^{\dagger}}             					
\newcommand{\nn}{\nonumber}							
\newcommand{\ee}{\mathrm{e}}             					
\newcommand{\ii}{\mathrm{i}}             					
\newcommand{\comm}[2]{\left[ #1, #2 \right]} 				
\newcommand{\diss}[1]{\mathcal{D}[ #1 ]}					
\newcommand{\op}[1]{\hat{#1}}							
\newcommand{\abs}[1]{\ensuremath{ \left| #1 \right| }}		
\newcommand{\abss}[1]{\ensuremath{ \left| #1 \right|^{2} }}	

\newcommand{\beq}{\begin{equation}}
\newcommand{\eeq}{\end{equation}}

\renewcommand{\H}[0]{\hat{H}}  						

\begin{document} 

\title{Quantum Zeno effect in the strong measurement regime of circuit quantum electrodynamics}
\author{D. H. Slichter}
\email[Electronic address: ]{slichter@berkeley.edu}
\altaffiliation[Present address: ]{Time and Frequency Division, National Institute of Standards and Technology, 325 Broadway, Boulder CO 80305}
\affiliation{Quantum Nanoelectronics Laboratory, Department of Physics, University of California, Berkeley CA 94720}

\author{C. M\"uller}
\affiliation{ARC Centre of Excellence for Engineered Quantum Systems, School of Mathematics and Physics, University of Queensland, Saint Lucia, Queensland 4072, Australia}
\affiliation{D\'epartement de Physique, Universit\'e de Sherbrooke, Sherbrooke, Qu\'ebec, Canada J1K 2R1}

\author{R. Vijay}
\altaffiliation[Present address: ]{Department of Condensed Matter Physics and Materials Science, Tata Institute of Fundamental Research, Mumbai, Homi Bhabha Road, Mumbai 400005, India}
\affiliation{Quantum Nanoelectronics Laboratory, Department of Physics, University of California, Berkeley CA 94720}

\author{S. J. Weber}
\altaffiliation[Present address: ]{MIT Lincoln Laboratory, 244 Wood Street, Lexington MA 02420}
\affiliation{Quantum Nanoelectronics Laboratory, Department of Physics, University of California, Berkeley CA 94720}

\author{A. Blais}
\affiliation{D\'epartement de Physique, Universit\'e de Sherbrooke, Sherbrooke, Qu\'ebec, Canada J1K 2R1}
\affiliation{Canadian Institute for Advanced Research, Toronto, Canada}

\author{I. Siddiqi}
\affiliation{Quantum Nanoelectronics Laboratory, Department of Physics, University of California, Berkeley CA 94720}

\begin{abstract}
We observe the quantum Zeno effect\textemdash where the act of measurement slows the rate of quantum state transitions\textemdash in a superconducting qubit using linear circuit quantum electrodynamics readout and a near-quantum-limited following amplifier.  Under simultaneous strong measurement and qubit drive, the qubit undergoes a series of quantum jumps between states.  These jumps are visible in the experimental measurement record and are analyzed using maximum likelihood estimation to determine qubit transition rates.  The observed rates agree with both analytical predictions and numerical simulations.  The analysis methods are suitable for processing general noisy random telegraph signals.  
\end{abstract}


\date{\today}

\maketitle

\section{Introduction}
The backaction of measurement is a peculiarly quantum mechanical phenomenon which gives rise to striking outcomes, such as the quantum Zeno effect (QZE).  In the QZE, the act of measurement inhibits transitions between eigenstates of the measured observable, slowing the state evolution of a ``watched'' quantum system.  The QZE was described in its modern form in 1977 by Misra and Sudarshan \cite{Misra1977}, although some related questions were tackled in prior papers \cite{Khalfin1958, Winter1961}.  The slowing of state evolution due to the QZE disappears in the classical limit $\hbar \rightarrow 0$, making the QZE a useful test for quantum behavior in a system \cite{Facchi2010, Bedingham2014}.  

The QZE was first observed experimentally in an ensemble of trapped ions \cite{Itano1990}, and has since been seen in a variety of other systems, including the electronic, nuclear, or motional states of atoms and molecules \cite{Nagels1997, Fischer2001, Streed2006, Balzer2002}, optical photons \cite{Kwiat1995, Kwiat1999, Hosten2006}, microwave photons \cite{Bernu2008}, and NV centers \cite{Wolters2013}.  In driven superconducting qubits, the QZE has been indirectly inferred from the transition between coherent Rabi oscillations and incoherent exponential population decay with increasing measurement strength \cite{Palacios-Laloy2010}, and by studying the dependence of this exponential decay on the time between discrete qubit projection pulses \cite{Kakuyanagi2015}. However, theoretical proposals also exist to observe the QZE in the quantum trajectory of a continuously monitored superconducting qubit \cite{Gambetta2008} or in the suppression by measurement of low-frequency superconducting flux qubit dephasing \cite{Matsuzaki2010}.

The QZE and related phenomena can have practical applications in quantum control and the engineering of decoherence.  Carefully designed measurements can be used to divide a larger Hilbert space into separate ``Zeno subspaces'' \cite{Facchi2002, Facchi2009}, where state evolution between subspaces is inhibited by the measurements.  The resulting ``quantum Zeno dynamics'' have recently been demonstrated experimentally \cite{Signoles2014, Schafer2014, Bretheau2015}.  This technique can even be used to generate multiparticle entanglement directly \cite{Wang2008}.  

In this work, we report the direct observation of the quantum Zeno effect in a superconducting qubit undergoing continuous strong measurement with simultaneous qubit driving.  The measurement record is analyzed to extract quantum jumps indicating individual qubit state transitions, and to determine the rates at which they occur in the presence of simultaneous qubit drive and measurement.  The extracted transition rates show inhibition of qubit state transitions due to the measurement, in agreement with both analytical Zeno theory and numerical simulations.  Additionally, we examine the excited state decay of the qubit during measurement, finding qualitative agreement with predictions for the Purcell decay of a multi-level qubit \cite{Boissonneault2010, Sete2014}.  The method for extracting transition rates can also be used to analyze general noisy random telegraph signals.  

\section{Theory of the Zeno effect}
In the absence of measurement, a resonantly driven qubit will undergo sinusoidal Rabi oscillations between states at frequency $\Omega$, where $\Omega$ depends on the strength of the resonant drive.  Repeated projective measurements made on this system at time intervals $\tau\ll 1/\Omega$ will tend to pin the qubit in one state, with occasional state transitions occurring as sudden quantum jumps.  The probability per unit time of a quantum jump out of the current state is given by (see e.g. Ref. \onlinecite{Cook1988}):
\beq
\label{disczenorate}
\Gamma=\frac{\Omega^2}{4 f_m},
\eeq
where $f_m=1/\tau$ is the frequency of the measurements.  The two major hallmarks of the QZE can be seen in this expression.  First, $\Gamma$ is independent of time, meaning that the state evolution under repeated measurement is exponential (linear in time at short times), whereas in the absence of measurement the qubit would exhibit sinusoidal state evolution (quadratic in time at short times). Qubit evolution processes which were already exponential in time, for example $T_1$ decay, remain exponential and are unaffected by the presence of repeated measurements.  Second, the transition rate varies inversely with the measurement frequency, tending to zero in the limit of infinitely frequent measurements.  However, the complete ``freezing'' of qubit evolution does not occur in real physical systems.  This is due to the energy-time uncertainty relationship, which causes the qubit to couple to an arbitrarily large energy spectrum (and thus arbitrarily many decay channels) as the time between measurements goes to zero \cite{Kofman2000}.

Our experiment uses a superconducting qubit in the circuit QED architecture \cite{Blais2004}. We bias the qubit in the dispersive regime, where the qubit-cavity detuning $\Delta$ is much larger than the qubit-cavity coupling $g$. In this limit, the cavity resonance frequency depends on the qubit state, enabling a quantum non-demolition measurement of the qubit state by driving the cavity near resonance and observing the response of its steady-state field. The measurement strength is directly related to the distinguishability of the different cavity field amplitudes conditioned on the qubit states.  For a linear cavity, the distinguishability is proportional to the amplitude of the measurement drive and could in principle be increased ad infinitum. In real systems, however, cavity non-linearities (both intrinsic and induced by coupling to the anharmonic qubit circuit) limit the obtainable field separations for readout, even at moderate drive amplitudes \cite{Slichter2012}.

Following Ref.~\cite{Gambetta2008}, and as shown in Appendix~\ref{analyticappend}, we derive the transition rate of a two-level qubit coupled to a cavity in the Zeno regime starting with the Jaynes-Cummings Hamiltonian in the dispersive regime with an additional qubit drive: 
\begin{align}
	\H = \omega_{r} a\hc a + \frac12 (\omega_{q}+\chi) \sigma_{z} + \chi a\hc a  \sigma_{z} + \frac12 ( \epsilon_{ro} a\hc \ee^{\ii \omega_{ro} t} +\epsilon_{ro}\cc a \ee^{-\ii \omega_{ro} t} ) + 
		\frac12 \left( \Omega  \sigma_{-} \ee^{-\ii \omega_d t} + \Omega \cc \sigma_{+} \ee^{\ii \omega_d t} \right) \,.
	\label{eq:H0}
\end{align}
Here, $a\; (a^\dagger)$ is the annihilation (creation) operator for photons in the cavity, the qubit is described by the Pauli matrices $\sigma$, and we have made a rotating wave approximation in the two driving terms.
The parameters are the cavity resonance frequency $\omega_{r}$, the qubit frequency $\omega_{q}$, the qubit-cavity coupling $g$, and the dispersive shift $\chi = g^{2}/\Delta$, where $\Delta=\omega_{r} - \omega_{q}$.
We apply a measurement drive to the cavity with frequency $\omega_{ro}$ and strength $\epsilon_{ro}$ and a Rabi drive to the qubit with frequency $\omega_d$ and amplitude $\Omega$.
We then perform a polaron transformation on the system, which effectively describes the cavity in a semi-classical picture with coherent state amplitudes that are conditioned on the qubit state. 
This transformation enables us to decouple the dynamical equations of the qubit and cavity and obtain a qubit-only reduced master equation, where the qubit dynamics depend mainly on qubit drive and measurement strength.
Because of the presence of the Rabi drive, this treatment is only valid in the case of small measurement strength, when the field amplitudes corresponding to the two qubit states are nearly indistinguishable.
Solving this master equation yields a qubit transition rate from the ground state to excited state in the presence of continuous circuit QED measurement~\cite{Gambetta2008}: 
\beq
	\label{contzenorate}
	\Gamma_{\uparrow,\mathrm{drive}}=\frac{\Omega^2}{2(\gamma_2+\Gamma_d)}\; ,
\eeq
where $\Gamma_d$ is the measurement-induced dephasing rate and $\gamma_2$ is the intrinsic qubit dephasing rate in the absence of measurement. 
The measurement rate is defined as $\Gamma_m=2 \Gamma_d$, as in Ref.~\onlinecite{Gambetta2008}.  In the strong measurement limit where $\Gamma_d \gg \gamma_2$, 
this expression is the same as for the discrete measurement case in Eq.~(\ref{disczenorate}) apart from a constant factor \cite{Schulman1998}.  
Although Eq.~(\ref{contzenorate}) is derived in the limit where the cavity linewidth $\kappa \gg \Gamma_m$, 
the similarity to Eq.~(\ref{disczenorate})\textemdash which has no such constraint in its derivation\textemdash suggests that the result may still be valid when $\kappa < \Gamma_m$, 
which is the regime of our experimental data.

\section{Experimental setup and calibrations}

\begin{figure}[tbp]
\begin{center}
\includegraphics[width=180mm]{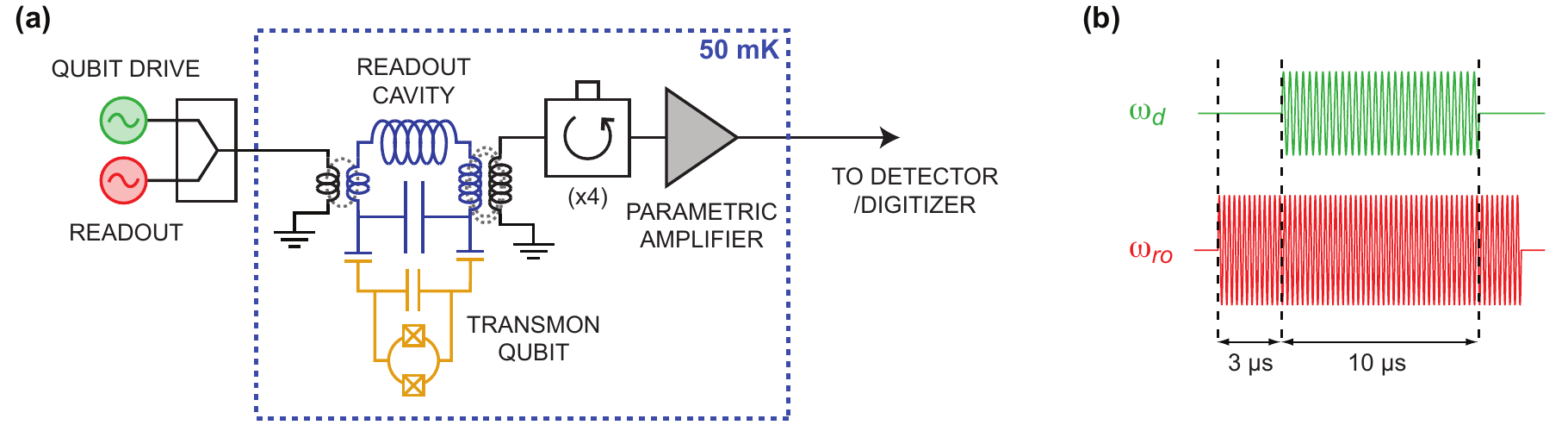}
\end{center}
\caption{\label{expsetup} Experimental setup.  A schematic of the experimental setup is shown in (a).  The readout and qubit drive tones are sent to the qubit/cavity circuit via a weakly coupled input port.  The readout signal from the cavity exits through the strongly coupled port and is amplified by a superconducting parametric amplifier, as well as further cryogenic and room temperature amplifiers (not shown), before being detected with homodyne mixing and digitized.  The Zeno experiment pulse sequence is shown in (b).  The readout is turned on 3 $\mu$s before the qubit drive to allow the cavity to come to steady state.  Only the data from the 10 $\mu$s with both qubit drive and readout are processed for Zeno rate analysis.  }
\end{figure}

The experimental setup is shown schematically in Figure \ref{expsetup}(a).  Our qubit is a planar two-junction transmon qubit~\cite{Koch2007}, tuned with a dc magnetic flux to a Lamb-shifted qubit frequency of $\tilde{\omega}_{q}/2\pi=5.3556$ GHz and an anharmonicity of $\alpha/2\pi=-258$ MHz.  We denote the two lowest transmon energy levels as $\ket{g}$ and $\ket{e}$, respectively.  The qubit is capacitively coupled ($g/2\pi=105.3$ MHz) to a lumped-element readout cavity with bare frequency of $\omega_r/2\pi=6.2724$ GHz and linewidth $\kappa/2\pi=7$ MHz.  Qubit measurement was performed by applying a readout drive to the cavity at $\omega_{ro}/2\pi=6.282$ GHz.  The readout signal from the cavity was amplified with a near-quantum-limited superconducting parametric amplifier~\cite{Vijay2011} followed by additional cryogenic and room temperature amplification stages, detected with homodyne mixing, and digitized at 10 ns intervals.  Both the qubit/cavity system and the parametric amplifier were anchored to the mixing chamber of a dilution refrigerator at 50 mK.

The average cavity photon occupation $\bar{n}$ and the measurement-induced dephasing rate $\Gamma_d$, which characterize the measurement strength, were determined by qubit spectroscopy with a simultaneous readout tone.  The value of $\bar{n}$ was determined from the ac Stark shift of the qubit frequency, accounting for qubit-induced cavity nonlinearities~\cite{Boissonneault2010}.  Because of the choice of readout frequency, $\bar{n}$ is different for the two qubit states $\ket{g}$ and $\ket{e}$; the ac Stark shift measurements give the average photon occupation with the qubit in the ground state, $\bar{n}_g$, since the qubit spectroscopy tone was much weaker than the saturation amplitude of the qubit transition. We use numerical simulations to infer the value of $\bar{n}_e$ as a function of the measured $\bar{n}_g$.  The value of $\Gamma_d$ is given by the half-width at half-maximum (HWHM) of the ac-Stark-shifted qubit line~\cite{Schuster2005}, which was determined by fitting to the measured line shape for several qubit drive powers and extrapolating to find the HWHM at zero qubit drive power.  This method allows us to calibrate out the effects of power broadening due to the qubit drive tone on the measured qubit linewidth.  With the readout tone off during qubit evolution, we measured a qubit relaxation time of $T_1=575$ ns and a pure dephasing time of $T_\varphi=7.9$ \us, corresponding to an intrinsic dephasing rate $\gamma_2=1/2T_1 + 1/T_\varphi=1$ \usinv.

The qubit drive frequency $\omega_d$ was chosen to be the ac-Stark-shifted qubit frequency $\tilde\omega_q(\bar{n}_g)$.  Since the qubit drive tone reaches the qubit via the cavity, its amplitude at the qubit (and thus the no-measurement Rabi frequency $\Omega$) depends on both its frequency and its amplitude at the cavity input.  We calibrated $\Omega$ as a function of resonant qubit drive amplitude over a range of qubit frequencies by tuning the qubit flux bias.  This allowed us to interpolate the value of $\Omega$ for a given $\omega_d$ and qubit drive amplitude.

Each iteration of the experiment consisted of a single measurement pulse lasting 17.5 $\mu$s, with simultaneous qubit drive applied for 10 $\mu$s beginning 3 $\mu$s after the measurement was turned on, as shown in Fig. \ref{expsetup}(b).  These times are chosen to be much longer than the relevant timescales $1/\kappa$, $1/\Gamma_m$, $1/\Omega$, and $T_1$, so that the experimental record captures the steady-state dynamics of the system.  We recorded $10^4$ such iterations, spaced 100 $\mu$s apart, for each combination of measurement strength and qubit drive strength.  The measurement record was digitized at $10^8$ samples/s.  

\section{Data processing and analysis}

The measurement records described above were analyzed to extract the transition rates from $\ket{g}\rightarrow\ket{e}$, denoted $\Gamma_\uparrow$, and from $\ket{e}\rightarrow\ket{g}$, denoted $\Gamma_\downarrow$.  
With multiple gigabytes of raw quantum jump data to analyze, the data processing algorithm must be able to operate with minimal user input and provide reliable output 
over a broad range of signal-to-noise ratio (SNR) and qubit transition rates, including low-SNR, high-rate scenarios.  

\begin{figure}[tbp]
\begin{center}
\includegraphics[width=7in]{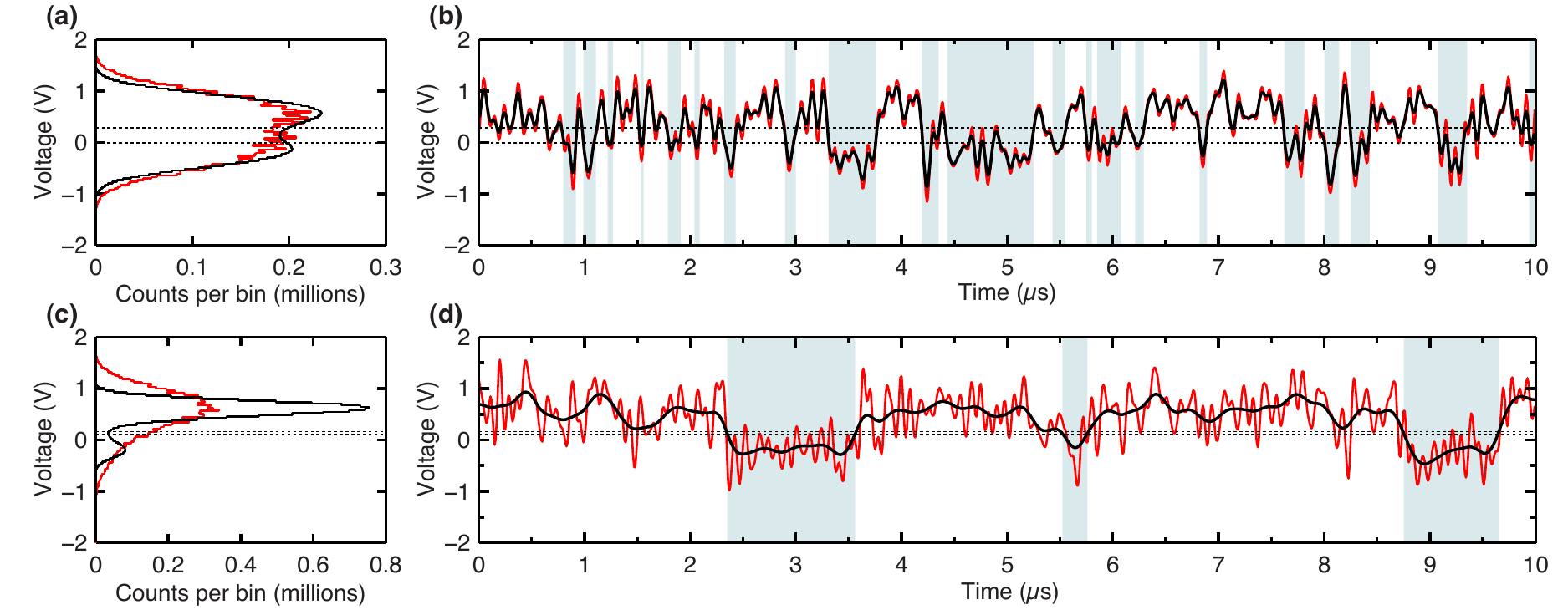}
\end{center}
\caption{\label{jumpfind} Data processing and quantum jump extraction.  Part (a) is a histogram of detected voltages from the full data set ($10^4$ readouts with $10^3$ points each) with $\Gamma_m=134$ \usinvsp ($\bar{n}_g=3.4$) and $\Omega/2\pi=3.6$ MHz, both for the raw data (red) and after filtering (black).  The hysteretic threshold voltages for state determination are shown as horizontal dashed lines.  In (b), a typical single measurement trace from the same data set is shown both as raw data (red) and after filtering (black), with the state determination threshold voltages shown as horizontal dashed lines.  The background color indicates the extracted qubit state: white for $\ket{g}$ and blue for $\ket{e}$.  Panels (c) and (d) show equivalent data for $\Gamma_m=134$ \usinvsp and $\Omega/2\pi=0.8$ MHz.  Because the transition rates are lower than in (a) and (b), the extraction algorithm performs heavier filtering, resulting in larger filtered SNR and smaller hysteresis between the threshold voltages.  }
\end{figure} 

We first perform filtering of the raw data to reduce the noise bandwidth and increase the SNR.  We use a zero-delay Gaussian finite-impulse-response (FIR) filter to ensure that state transition edges will remain smooth and will not be shifted in time by the filtering process.  The order of the filter must be chosen to reduce the noise without degrading the desired signal, and so will depend both on the initial SNR and the transition rate between states.  For each experimental bias point, we histogram all $10^7$ data points taken during simultaneous qubit driving and readout ($10^4$ traces of $10^3$ points each).  Example histograms of raw data are shown as the red curves in Fig. \ref{jumpfind}(a) and (c).  We then perform filtering and histogram the filtered data points, shown as black curves in Fig. \ref{jumpfind}(a) and (c).  The histogram of the filtered data forms a bimodal distribution~\footnote{If the histogram is not bimodal, as for the raw data in Fig. \ref{jumpfind}(c), we perform additional filtering until it becomes bimodal.}, with a minimum between the two peaks corresponding to the two qubit states.  The height of this minimum is dictated both by the amount of noise on the signal and the number of transitions in the measurement record.  Filtering will reduce the noise amplitude, and thus the width and overlap of the bimodal peaks, leading to a reduction in the height of the minimum.  However, filtering will also slow the rise and fall times of state transition edges, which will increase the height of the minimum between histogram peaks.  Based on these considerations, we designate the ``optimal'' filter order as the one which gives the lowest minimum between the bimodal peaks of the resulting voltage histogram, and determine it by applying filters of increasing order to the raw data until the height of the minimum between the peaks is as small as possible.  This provides a simple, robust method for choosing filter order.  The ``optimal'' filter order was determined separately for each combination of qubit drive strength and measurement strength.  

After the data have been filtered, we determine the qubit state at each time point using a thresholding algorithm.  For increased robustness to noise, we use two hysteretic thresholds with ``Schmitt trigger'' behavior~\cite{Yuzhelevski2000}; state transitions are registered when the voltage crosses the higher threshold going upwards (if the state was low), or when it crosses the lower threshold going downwards (if the state was high).  The voltage thresholds are shown as dotted horizontal lines in Fig. \ref{jumpfind}.  Figures \ref{jumpfind}(b) and (d) show representative individual data traces from the data sets histogrammed in Fig. \ref{jumpfind}(a) and (c), respectively, corresponding to two different values of $\Omega$ with identical $\Gamma_m$.  The red curves show the raw data, while the black curves are after filtering.  The blue and white background colors show the extracted qubit state ($\ket{e}$ and $\ket{g}$, respectively) as determined by the thresholding algorithm.  The data in (c) and (d) have fewer state transitions, and thus heavier filtering can be employed to increase the SNR.  

The location of the thresholds is determined using histograms of the filtered voltages, and depends on the voltage $V_m$ of the minimum between the histogram peaks and the voltages $V_h$ ($V_l$) and the HWHM $w_h$ ($w_l$) of the high (low) histogram peaks.  The value of $w_h$ ($w_l$) is derived from the half-height of the histogram above (below) the corresponding peak $V_h$ ($V_l$), a method that works even when the histogram peaks are not well-separated.  We define the SNR of the filtered data to be $\sqrt{2 \,\mathrm{ln}\, 2} (V_h-V_l)/(w_h+w_l)$.  The thresholds are chosen to be $V_m + \frac{w_h^2}{2\, \mathrm{ln}\, 2(V_h-V_l)}$ and $V_m - \frac{w_l^2}{2\, \mathrm{ln}\, 2(V_h-V_l)}$, as suggested in Ref.~\cite{Yuzhelevski2000}, although unlike that work we do not perform iterative refinement of the thresholds.  The distance between the thresholds increases as the SNR decreases to reduce the likelihood of spurious state transitions being registered due to noise.  For each measurement trace, we determine the initial state by comparing the first data point with $V_m$.  Given a low (high) state, we look for the first upward crossing of the higher threshold (downward crossing of the lower threshold) and note the dwell time before the state change.  We continue in this manner, alternating thresholds, until the end of the trace is reached, noting down all dwell times in both the high and low states.  For each measurement trace, exactly one dwell time is cut short by the end of the trace and is marked as ``right-censored''~\cite{Klein2003}.  

\begin{figure}[tbp]
\begin{center}
\includegraphics[width=7in]{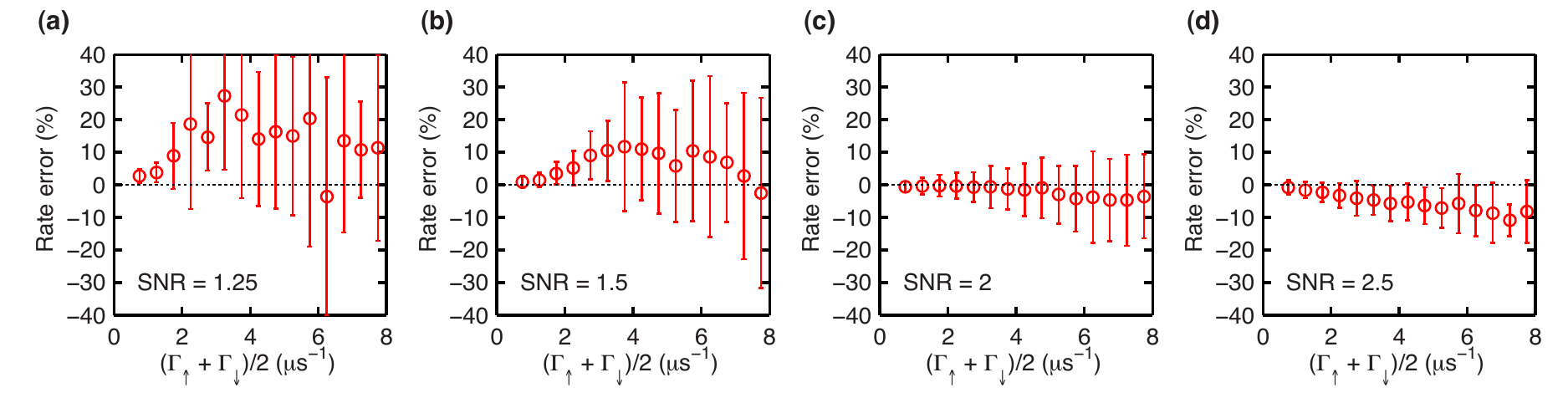}
\end{center}
\caption{\label{extraction} Performance of rate extraction algorithm.  We show the percentage deviation of the extracted mean transition rate $(\Gamma_\uparrow + \Gamma_\downarrow)/2$ from the true mean transition rate as a function of the extracted mean rate, as determined from roughly 7,000 simulated data sets.   Error bars show the root mean squared error.  The four panels correspond to four different values of the initial SNR, as described in the text.}
\end{figure}

We use maximum likelihood estimation to determine the transition rates between states given the set of observed dwell times in each state.  However, the filtering can cause some fast voltage excursions in the raw data, like the one in Fig. \ref{jumpfind}(d) near 8.25 \us, not to register as state transitions.  In general, the finite bandwidth of the measurement chain, combined with filtering described above, skews the distribution of observed dwell times toward longer times.  We compensate for this in our analysis by assuming a probability distribution for the dwell times which takes these effects into account~\cite{Naaman2006}.  For right-censored dwell times, the observed time represents a lower bound on the dwell time, rather than an exact value.  However, we can include this partial information in the likelihood function as well, which is particularly useful for data sets with relatively few state transitions and thus many censored dwell times.  We emphasize that this maximum likelihood method can be applied to dwell times extracted using any technique; for example, it can be used with dwell times determined from wavelet analysis, which is more robust than thresholding for signals with substantial low-frequency drifts~\cite{Prance2015}.  Details on the maximum likelihood estimation and functional forms of the likelihood function are provided in Appendix \ref{mleappend}.

The extraction algorithm was tested on simulated noisy random telegraph data with a variety of experimentally relevant transition rates and signal-to-noise ratios (SNRs).  Note that these data are not the same as the numerical simulations of the qubit dynamics described in the next section.  The 3 dB noise bandwidth (14 MHz) and sample rate ($10^8$ sample/s) were the same as in the raw experimental data.  The error in the extracted mean transition rate $(\Gamma_\uparrow + \Gamma_\downarrow)/2$, relative to the true mean transition rate, is plotted in Fig. \ref{extraction} for four values of the initial SNR before filtering.  For simulated data at the highest rates shown, the mean dwell time was 130 ns, while the shortest oscillation timescale of the simulated noise was around 70 ns, making the task of distinguishing signal from noise challenging.  Despite this, the mean systematic bias in the extracted rates is below 12\%, even at an SNR of only 1.5.  This performance is also notable because the algorithm operates autonomously, without the need for externally provided guesses or input, across the entire range of transition rates and SNR shown.  

The transition rates extracted from the simulated noisy random telegraph data were used to calibrate systematic bias and systematic uncertainty in the rates extracted from the experimental data, by matching experimental traces to simulated data sets with similar extracted transition rates and filtered SNR.  The median systematic bias and median systematic uncertainty for the experimental data were -3.5\% and 2\% of the extracted rates, respectively.  The magnitude of the systematic bias (uncertainty) was below 20\% (10\%) of the extracted rate for all experimental data points, and below 9\% (5\%) of the extracted rate for 90\% of the experimental data points.

\section{Numerical simulations}

We also performed numerical simulations of the qubit/cavity system over a range of bias points, using the generalized Jaynes-Cummings Model for a multi-level qubit coupled to a cavity.  
We simulated the dynamics numerically, including the measurement, with a stochastic master equation~\cite{Wiseman:1993a, Gambetta2008, Wiseman:book}, using experimentally measured parameters for the qubit and cavity. 
The master equation included qubit relaxation and dephasing as well as cavity photon loss. 
The measurement and qubit drive were simulated as two independent coherent microwave drives acting on the cavity.  
The simulations yield measurement records for direct comparison with experiment, as well as records of qubit populations, complex cavity field amplitudes,  and cavity photon occupation numbers. 
Details are given in Appendix \ref{numappend}.  Qubit transition rates were then determined from the simulated qubit state population record using the same rate extraction algorithm used to process the experimental data.  

The experimental data, and thus the rate extraction algorithm, do not distinguish between the qubit state $\ket{e}$ and higher excited states.  This occurs because we operate the parametric amplifier in phase-sensitive mode to achieve the lowest noise performance, and thus only one quadrature of the readout signal from the cavity is amplified~\cite{Vijay2011}.  The phase of the amplified quadrature was chosen to maximize the ability to discriminate between states $\ket{g}$ and $\ket{e}$; however, because of the parameters of the qubit/cavity system, the projection of the complex cavity amplitudes corresponding to higher excited states onto this choice of amplified quadrature was essentially the same as that for $\ket{e}$, rendering them indistinguishable from $\ket{e}$ in the measurement record~\cite{Slichter2011}.  The populations of higher excited states were measured to be $\sim1-5$\% using a different data set which distinguished $\ket{e}$ from higher qubit states by choosing a different amplified quadrature, at the expense of substantially decreased SNR.
These values are corroborated by the numerical simulations.  However, the presence of qubit population outside the $\{\ket{g},\ket{e}\}$ manifold is not expected to have an effect on the estimates of $\Gamma_{\uparrow}$.

\section{Results}

We first examine the qubit excitation rate during simultaneous measurement and qubit driving to look for the QZE.  
Since $\bar{n}_g\neq\bar{n}_e$ for our experimental parameters, the qubit drive tone at the ac-Stark-shifted qubit frequency $\tilde\omega_q(\bar{n}_g)$ was resonant when the qubit was in the ground state, but not when it was in the excited state.  For this reason, our analysis of driven transition rates is restricted to the $\ket{g}\rightarrow\ket{e}$ transition rate, denoted $\Gamma_\uparrow$, which can be expressed as the sum of three rates:
\beq
\label{totalrate}
\Gamma_\uparrow=\Gamma_{\uparrow,\mathrm{drive}}+\Gamma_{\uparrow,\mathrm{DD}}+\Gamma_{\uparrow,\mathrm{th}}.
\eeq
Here $\Gamma_{\uparrow,\mathrm{drive}}$ is the transition rate due to the qubit drive, $\Gamma_{\uparrow,\mathrm{DD}}$ is the contribution from dressed dephasing \cite{Boissonneault2009, Slichter2012}, and $\Gamma_{\uparrow,\mathrm{th}}$ represents thermal excitation of the qubit.  To isolate $\Gamma_{\uparrow,\mathrm{drive}}$ for comparison to the predictions in Eq. (\ref{contzenorate}), we determine $\Gamma_\uparrow$ as a function of measurement strength in the absence of qubit drive ($\Omega=0$), which is equal to $\Gamma_{\uparrow,\mathrm{DD}}+\Gamma_{\uparrow,\mathrm{th}}$.  We can then subtract this contribution from the total rate $\Gamma_\uparrow$ in the presence of qubit drive to find $\Gamma_{\uparrow,\mathrm{drive}}$.  This calibration was performed for both experimental data and numerically simulated data.  The value of $\Gamma_{\uparrow,\mathrm{DD}}+\Gamma_{\uparrow,\mathrm{th}}$ was measured to be between 0.018 \usinvsp and 0.022 \usinvsp for all bias points, which is considerably smaller than $\Gamma_{\uparrow,\mathrm{drive}}$ for all but the lowest values of $\Omega$.  

\begin{figure}[t]
\begin{center}
\includegraphics[width=7in]{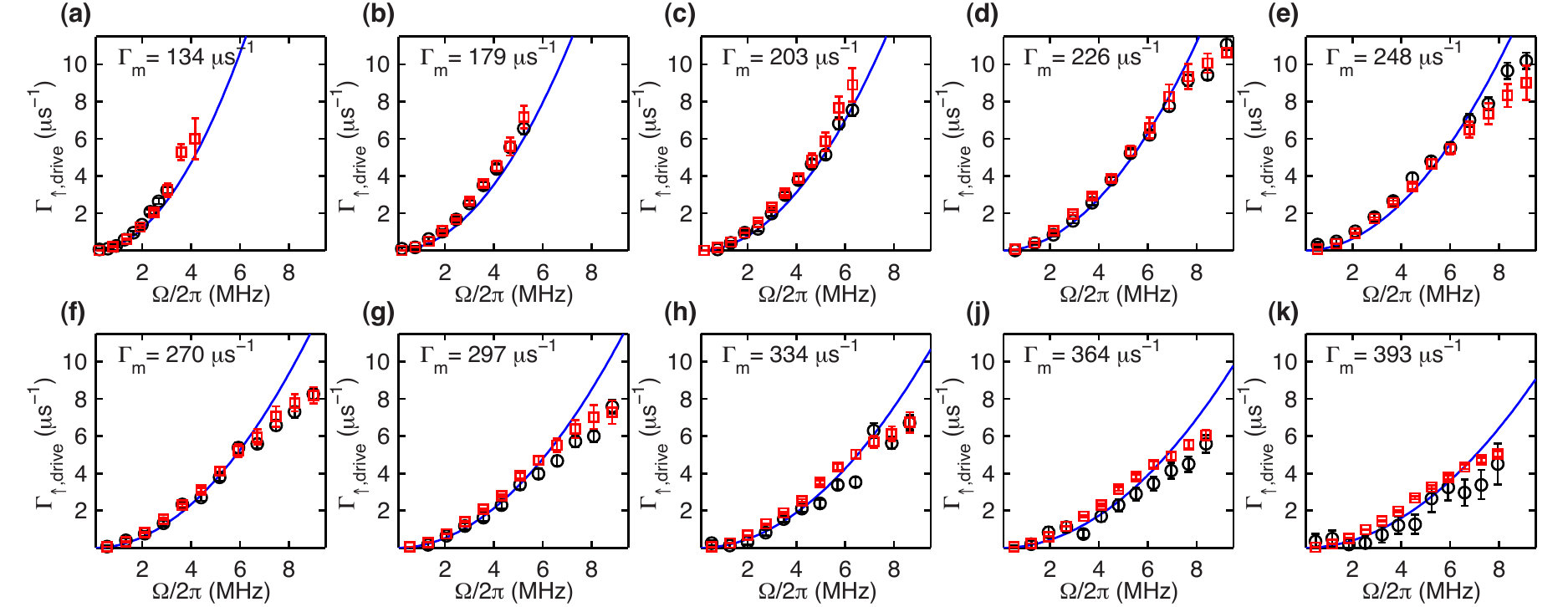}
\end{center}
\caption{\label{gamuplinecut} Driven transition rates during measurement.  We plot values of $\Gamma_{\uparrow,\mathrm{drive}}$ versus qubit drive strength $\Omega$ for ten measurement strengths $\Gamma_m$ ranging between 134 \usinvsp ($\bar{n}_g=3.4$) and 393 \usinvsp ($\bar{n}_g=37$).  Experimental data (red squares) are shown along with numerically simulated data (black circles) and theoretical values calculated from Eq. (\ref{contzenorate}) with no adjustable parameters (blue lines).  Error bars represent 95\% confidence intervals.}
\end{figure}

\begin{figure}[t]
\begin{center}
\includegraphics[width=7in]{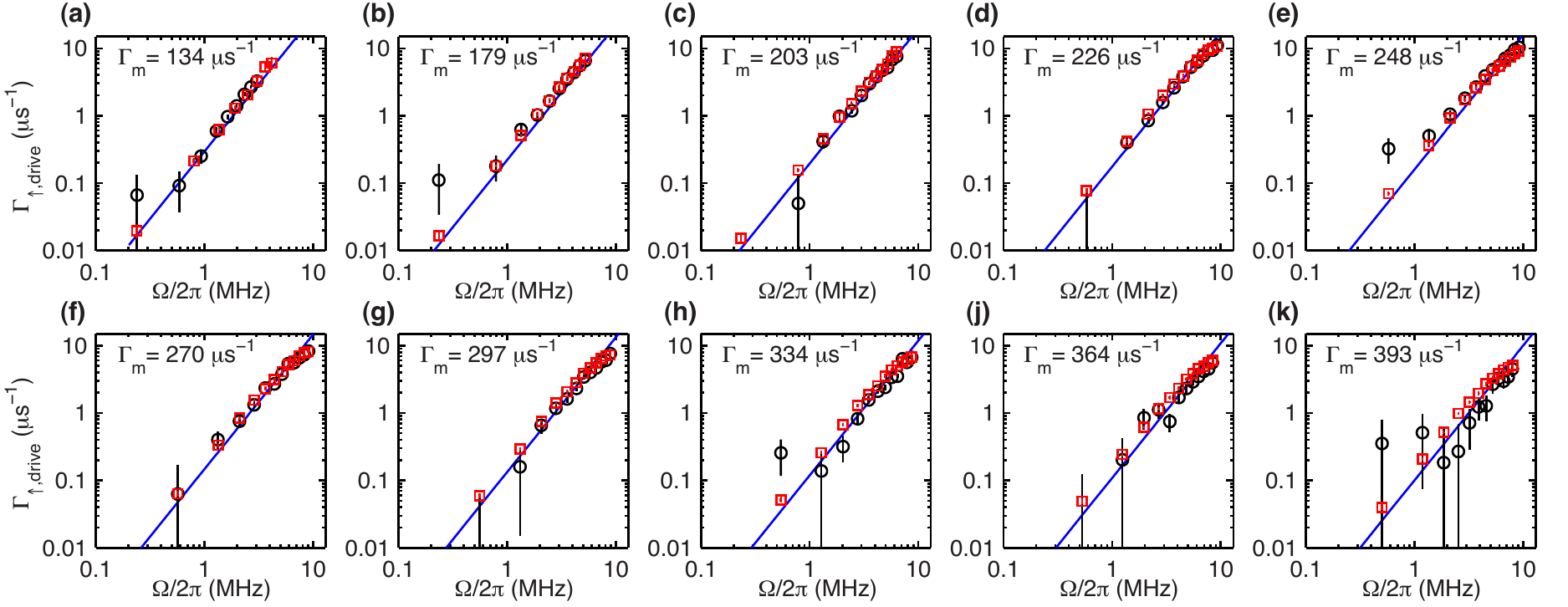}
\end{center}
\caption{\label{gamuplinecutloglog} Driven transition rates during measurement on logarithmic axes. The data are the same as in Fig. \ref{gamuplinecut}, but the logarithmic axes highlight the agreement between experiment, analytical theory, and numerics for the smallest $\Omega$, as well as the $\Omega^2$ scaling expected from Eq.~\eqref{contzenorate}.  Experimental data (red squares) are shown along with numerically simulated data (black circles) and theoretical values calculated from Eq. (\ref{contzenorate}) with no adjustable parameters (blue lines).  Error bars represent 95\% confidence intervals.
}
\end{figure}

Figures~\ref{gamuplinecut} and \ref{gamuplinecutloglog} show the extracted $\Gamma_{\uparrow,\mathrm{drive}}$ as a function of $\Omega$ for ten different measurement strengths ranging from $\Gamma_m=134$ \usinvsp to $\Gamma_m=393$ \usinv, corresponding to values of $\bar{n}_g$ between 3.4 and 37.  The rates from the experimental data and the numerical simulations are plotted as red squares and black circles, respectively, while the rates predicted by the analytical theory in Eq. (\ref{contzenorate}), with no adjustable parameters, are plotted as solid blue lines.  

The experimental data and numerics are in good agreement, and both coincide with the prediction of the analytical theory at most points, showing the presence of the quantum Zeno effect across a broad range of measurement strengths and qubit drive amplitudes.  Some deviations can be seen above $\Omega/2\pi \approx 6$ MHz, where both the experiment and numerics give consistently lower values of $\Gamma_{\uparrow,\mathrm{drive}}$ than predicted by the analytical theory.  Because the numerical simulations exhibit the same behavior as the experiment, we believe this is a real effect and not an experimental artifact.  We postulate that this additional slowing of the transition rate may be due to higher order terms neglected in the derivation of the analytical theory under the approximation $\Gamma_m \ll \kappa$, since this approximation is no longer valid here.  The effect is somewhat more pronounced for larger $\Gamma_m$.  This could also be associated with the breakdown of the dispersive approximation for $\bar{n}_g>n_\mathrm{crit}=\Delta^2/4g^2$, which corresponds in our system to $\Gamma_m\approx340$ \usinv.

We additionally examine the qubit $T_1$\textemdash  the time constant for transitions from $\ket{e}$ to $\ket{g}$\textemdash with measurement on but no qubit drive.  In the simple two-level picture used to derive Eq.~(\ref{disczenorate}), transition rates from exponential processes such as $T_1$ decay are unchanged by the QZE.  However, in our driven multilevel system other effects are expected to give some dependence of $T_1$ on measurement strength.  
In circuit QED, the $T_1$ decay rate can in general be parameterized as the sum of the Purcell decay rate $\Gamma_\mathrm{P}$ and a decay rate $\Gamma_\mathrm{NR}$ from nonradiative loss channels such as dielectric loss \cite{Houck2008, Reed2010}.  
The Purcell effect can be thought of as the decay of the photonic part of the qubit eigenstate in the coupled qubit-cavity system, as seen in the expression for the Purcell rate~\cite{Boissonneault2010,Sete2014}
\beq
\label{purc}
\Gamma_\mathrm{P}=\kappa \left |\overline{\bra{g,n}}a\overline{\ket{e,n}}\right |^2 \,,
\eeq
where $\overline{\ket{g,n}}$ and $\overline{\ket{e,n}}$ are eigenstates of the generalized Jaynes-Cummings Hamiltonian for a multilevel qubit corresponding most closely to $n$ photons in the bare cavity and the bare qubit in $\ket{g}$ or $\ket{e}$, respectively. 
The matrix elements in Eq.~\eqref{purc} can be calculated numerically, and become smaller for increasing $n$.  As a result, the Purcell decay of the qubit is suppressed by the presence of photons in the readout cavity \cite{Boissonneault2010, Sete2014}.  Note that in our case we replace $n$ with \nbare, because Purcell decay from $\ket{e}$ occurs with \nbare~photons in the cavity initially.  Details of the calculation are given in Appendix~\ref{purcappend}.

\begin{figure}[tbp]
\begin{center}
\includegraphics[width=5.5in]{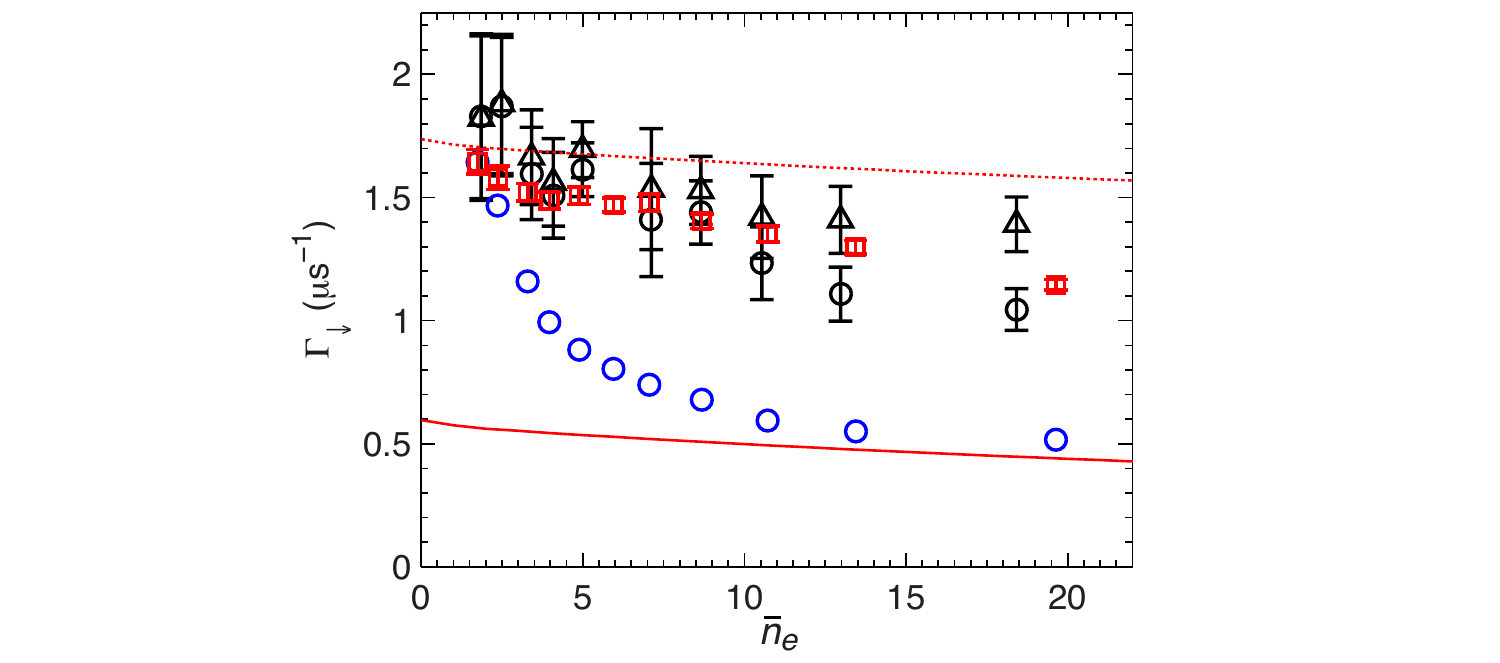}
\end{center}
\caption{\label{t1} $\Gamma_\downarrow$ versus $\bar{n}_e$ with no qubit drive.  The red squares are measured values of the qubit decay rate as a function of $\bar{n}_e$.  The black triangles and circles are values derived from numerical simulations; the former distinguishes between state $\ket{e}$ and higher states in the rate extraction process, while the latter does not (as is the case for our experimental data).  These values differ somewhat for $\bar{n}_e\gtrsim 8$, indicating the presence of population in higher excited states.  The solid red line is the predicted Purcell decay rate $\Gamma_\mathrm{P}$ from Eq. (\ref{purc}), and the dashed red line is $\Gamma_\mathrm{P} + \Gamma_\mathrm{NR}$ as described in the text.  The blue circles show scaling of the qubit decay rate that would be expected from Eq. (\ref{contzenorate}) if the decay were subject to the QZE.  Error bars represent 95\% confidence intervals.}
\end{figure}

Figure \ref{t1} shows the measured decay rate as a function of \nbare~(red squares).  The $T_1$ decay of the qubit is suppressed in the presence of readout photons; however, 
the overall decay rate is higher than $\Gamma_\mathrm{P}$ (red solid line) for all \nbare, indicating that our experiment is not limited by Purcell decay and that $\Gamma_\mathrm{NR} > 0$.  
We derive a value for $\Gamma_\mathrm{NR}$ by subtracting the zero-photon Purcell decay rate $\kappa g^2/\Delta^2$ from the experimental value of $\gamma_1=1/T_1$ measured with no readout photons present ($\bar n_{e} = 0$) and plot the theoretically predicted total decay rate $\Gamma_\mathrm{P} + \Gamma_\mathrm{NR}$ as the red dotted line.  This prediction is reasonably close to the experimental data for small $\bar{n}_e$, but deviates substantially with increasing $\bar{n}_e$.  To determine whether this is an experimental artifact (due to readout-power-dependent dielectric loss in the transmon capacitor~\cite{OConnell2008}, for example), we also performed numerical simulations of the qubit decay with no Rabi drive.  The resulting rates (black circles and black triangles) are also consistently lower than the predictions of Eq. (\ref{purc}) and in reasonable agreement with the experimental data.  The numerics also allow us to determine the effect of qubit population in states higher than $\ket{e}$ on the observed decay rate; the black circles show the observed $T_1$ decay rate when $\ket{e}$ is not distinguished from higher qubit states, as is the case with our experimental data, while the black triangles show the true $T_1$ decay rate from $\ket{e}\rightarrow\ket{g}$ when $\ket{e}$ is distinguished from higher qubit states.  The difference between these traces becomes significant for $\bar{n}_e\gtrsim 8$ but does not appear large enough to fully account for the discrepancy between the experimental data and the analytical theory.  We believe that other effects, such as dressed dephasing \cite{Boissonneault2009, Slichter2012} and qubit-induced non-linearities, are responsible for this deviation from the simple theory in Eq. (\ref{purc}).  The blue circles shows how the qubit decay rate would depend on $\bar{n}_e$ if it were subject to the QZE and scaled as $1/\Gamma_m$, as in Eq. (\ref{contzenorate}), relative to the measured value for the lowest \nbare.  The large disparity between these points and the experimental and numerical data indicate that the variation of $\Gamma_\downarrow$ with $\bar{n}_e$ is not due to the QZE; this agrees with the expected result that the QZE does not affect exponential processes such as $T_1$ decay.


\section{Conclusions}

We have observed the quantum Zeno effect in a continuously measured superconducting qubit and demonstrated quantitative agreement with both analytical theory and numerical simulations.  Interestingly, the agreement holds even for measurement strengths much larger than are allowed by the assumptions of the analytical derivation.  We have shown that $T_1$ decay is not subject to the quantum Zeno effect, as predicted.  We do observe an gradual increase in the qubit $T_1$ with increasing measurement strength, in qualitative agreement with the simple picture from the Purcell effect in circuit QED.  We have also demonstrated a robust algorithm for determining transition rates from our measurement record, which can be readily applied to the analysis of general noisy random telegraph signals.

\begin{acknowledgments}
We thank J. Gambetta, O. Naaman, J. Aumentado, and Y. C. Hagar for useful discussions and K. Lalumi\`ere for help with the stochastic master equation simulations. D.H.S. acknowledges support from a Hertz Foundation Fellowship endowed by Big George Ventures.  R.V. and I.S. acknowledge support from the ARO QCT program.  A.B. acknowledges support from NSERC.  We thank Calcul Qu\'ebec and Compute Canada for computational resources.

D.H.S. and R.V. designed and carried out the experiments.  S.J.W. fabricated the qubit sample.  D.H.S. analyzed the data and developed the rate extraction algorithm.  C.M. performed the theoretical calculations and numerical simulations.  A.B. and I.S. supervised the work.  D.H.S. and C.M. wrote the manuscript with input from all authors.  
\end{acknowledgments}

\appendix

\section{Details of analytical Zeno derivation \label{analyticappend}}

Here we derive the expression for qubit transition rates given in Eq. \eqref{contzenorate}.   We will focus only on the steps necessary to derive the qubit transition rates; more detail can be found in Ref.~\onlinecite{Gambetta2008}.  We note that this derivation is for the case of a two-level qubit, not a multi-level qubit such as a transmon.  
We describe the qubit-cavity system in the Jaynes-Cummings model (setting $\hbar=1$):
\begin{align}
	\H_{JC} = \omega_{r} a\hc a +\frac12 \omega_{q} \sigma_{z} + g (a\hc \sigma_{-} + a \sigma_{+}) + \frac12 (\epsilon_{ro} \ee^{-\ii\omega_{ro}t} a\hc + \epsilon_{ro}\cc \ee^{\ii\omega_{ro}t} a),
\end{align}
where $\omega_{r}$ is the cavity frequency, $\omega_{q}$ the qubit energy, $g$ is the qubit-cavity coupling, and $\epsilon_{ro}$ is the amplitude of the cavity drive.
Here $a$ is the annihilation operator for cavity photons, and the Pauli matrices $\sigma$ describe the qubit.  For large qubit-cavity detuning, such that $\Delta = \omega_{r} - \omega_{q} \gg g$, we move into the dispersive frame \cite{Blais2004}
\begin{align}
	\H_{disp} = (\omega_{r} - \omega_{ro}) a\hc a + \frac12 \tilde\omega_{q} \sigma_{z} + \chi a\hc a \sigma_{z} + \frac12 ( \epsilon_{ro} a\hc +\epsilon_{ro}\cc a ),
	\label{eq:HDisp}
\end{align}
where $\chi = g^{2}/\Delta$ is the dispersive shift and $\tilde\omega_{q} = \omega_{q} + \chi$ is the Lamb-shifted qubit frequency. We also move the cavity into a rotating frame at frequency $\omega_{ro}$ and assume that $\omega_{ro}$ is far detuned from the qubit transition $\omega_{q}$.  For a cavity drive used for a dispersive measurement, this condition is fulfilled. Since we are concerned with the qubit's dynamics under simultaneous qubit drive and measurement, we add a qubit driving term at frequency $\omega_d$ to Eq. \eqref{eq:HDisp}, writing
\begin{align}
	\H_{qd} = \frac12 \left( \Omega \sigma_{-} \ee^{-\ii \omega_d t} + \Omega\cc \sigma_{+} \ee^{\ii \omega_d t} \right),
\end{align}
where $\Omega$ is the qubit Rabi frequency.  Hereafter we examine the case of resonant qubit drive, where $\omega_d = \tilde\omega_{q}$. In a frame rotating at the qubit drive frequency, the uncoupled qubit dynamics are described by $\H_{q} = \frac12 \Omega \sigma_{x}$, 
leading to coherent qubit flopping at the Rabi frequency. The complete dynamics of the system will be described by a master equation of the form
\begin{align*}
	\dot \rho = -\ii \comm{\H_{disp} + \H_{qd}}{\rho} + \kappa \diss{a}\rho + \gamma_{1} \diss{\sigma_{-}} \rho + \gamma_{\varphi} \diss{\sigma_{z}}\rho \, ,
\end{align*}
with the cavity loss-rate $\kappa$, the qubit decay rate $\gamma_{1}$, and the qubit pure dephasing rate $\gamma_{\varphi}$.  The total qubit dephasing rate is then $\gamma_{2} = \frac12 \gamma_{1} + \gamma_{\varphi}$.
The dissipative superoperators are defined by $\diss{\op o}\rho = \op o \rho \op o\hc -\frac12 (\op o\hc \op o \rho + \rho \op o\hc \op o)$.

Next we apply a polaron transformation to the system, defined by~\cite{Gambetta2008}
\begin{align}
	P = \sum_{i} D(\alpha_{i}) \ket i \bra i \, ,
\end{align}
where $D(\alpha) = \exp\{ \alpha a\hc - \alpha\cc a \}$ is the usual field displacement operator and the sum is over the qubit states. 
The polaron transformation applies a qubit-state-dependent shift to the cavity field, where the complex-valued displacement amplitudes $\alpha_{i}$ 
are the cavity field amplitudes conditioned on the qubit states and are in general time-dependent. 
The equations of motion for the $\alpha_{i}$ follow from the polaron transformation under the constraint that the resulting Hamiltonian represent an un-driven cavity. 
This treatment and the resulting equations for the cavity field are similar to a semi-classical treatment of the cavity dynamics, while still keeping track of intrinsic and qubit-induced non-linearities~\cite{Boissonneault:2012}.
The polaron frame best captures the dynamics of the full system when the cavity field can be accurately described as a coherent state whose complex amplitude depends on the instantaneous qubit state.
In practice the use of the polaron frame is restricted to weak measurement regime, as becomes evident when considering its effect on the qubit ladder operators:
\begin{align}
	P\hc \sigma_{-} P = D_{\beta} \sigma_{-} \, . 
	\label{eq:SigmaPol}
\end{align}
Here the generalized displacement operator $D_{\beta}$ is defined by
\begin{align}
	D_{\beta} = D\hc(\alpha_{0}) D(\alpha_{1}) = D(\beta) \ee^{-\ii \varphi} \, ,
\end{align}
with the measurement separability $\beta = \alpha_{1} - \alpha_{0}$ and the phase $\varphi = \text{Im}(\alpha_{0} \alpha_{1}\cc)$. This follows naturally from the fact that the polaron transformation connects a given qubit state with a corresponding coherent state in the cavity. 
Any change in the qubit state must be connected with a simultaneous change in the cavity field.
Importantly, the transformation Eq. \eqref{eq:SigmaPol} results in an expression that, for non-vanishing values of $\abs\beta$, contains all orders of the cavity annihilation and creation operators.  This makes it impractical to find a reduced description of the dynamics of the qubit alone, 
except for the weak measurement case, when $\abs\beta \ll 1$ and $D_{\beta} \sim \mathds{1}$.
In this case the system reduces to a set of equations for the classical field amplitudes $\alpha_{i}$ conditioned on the qubit state 
and a Hamiltonian describing the qubit coupled to the $\alpha_{i}$ and an effectively un-driven cavity, describing the quantum fluctuations of the cavity field.

By applying the polaron transformation to the dissipators, we immediately find the measurement-induced dephasing since 
\begin{align*}
	\kappa \diss{a}\rho \rightarrow \kappa\diss{a}\rho + \Gamma_{d} \diss{\sigma_{z}}\rho + \ldots \, ,
\end{align*}
where $\Gamma_{d} = \frac12 \kappa \abss\beta$ is the measurement-induced dephasing rate and the omitted terms can be absorbed into the coherent part of the master equation.

We now find the qubit-only dynamics by tracing the master equation over the cavity degrees of freedom in the Fock basis $\ket n$, to find the time evolution of the qubit's reduced density matrix:
\begin{align}
	\tilde\rho = \sum_{n} \bra n \rho \ket n\,.
\end{align}
Finally, the jump rate in Eq. \eqref{contzenorate} can be found from the qubit's Bloch vector elements $\mean{\sigma_{i}}$. 
For resonant qubit drive, where $\omega_d = \tilde\omega_{q}$, one finds
\begin{align}
	\dot{\mean{\sigma_{x}}} &= - (\gamma_2+\Gamma_d) \mean{\sigma_{x}} \,,\nn\\
	\dot{\mean{\sigma_{y}}} &= - (\gamma_2+\Gamma_d) \mean{\sigma_{y}} + \Omega \mean{\sigma_{z}} \,,\nn\\
	\dot{\mean{\sigma_{z}}} &= \Omega \mean{\sigma_{y}} - \Gamma_{1} \mean{\sigma_{z}} + \l( \gamma_{\uparrow} - \gamma_{\downarrow} \r)\,,
\end{align}
where $\Gamma_{1} = \gamma_{\uparrow} + \gamma_{\downarrow}$ includes all dissipative transitions between qubit levels (i.e. thermally induced by the environment as well as dressed dephasing).
Assuming that the measurement-induced dephasing dominates, such that $\dot{\mean{\sigma_{x}}} = \dot{\mean{\sigma_{y}}} = 0$, the equation of motion for $\mean{\sigma_{z}}$ takes the form:
\begin{align}
	\dot{\mean{\sigma_{z}}} &= - \l( \frac{\Omega^{2}}{\gamma_2+\Gamma_d} + \Gamma_{1} \r) \mean{\sigma_{z}} + \l( \gamma_{\uparrow} - \gamma_{\downarrow} \r) \,,\nn\\
		&= - \l( \Gamma_{\uparrow} + \Gamma_{\downarrow} \r) \mean{\sigma_{z}} + \l( \gamma_{\uparrow} - \gamma_{\downarrow} \r)
\end{align}
We identify $\Gamma_{\uparrow} = \Gamma_{\uparrow,\text{drive}} + \gamma_{\uparrow}$ and $\Gamma_{\downarrow} = \Gamma_{\downarrow,\text{drive}} + \gamma_{\downarrow}$ and find the drive-induced transition rate as in Eq. (\ref{contzenorate}):
\begin{align}
\Gamma_{\uparrow, \text{drive}} = \Gamma_{\downarrow, \text{drive}} =\frac{\Omega^{2}}{2(\gamma_2+\Gamma_d)}.
\end{align}

\section{Details of the numerical simulations \label{numappend}}
	
	In contrast to the simple two-level qubit model used in the previous description, the transmon qubit used in our experiments is a genuine multi-level system. In the presence of strong driving and measurement, which is the regime of our experiments, the additional levels lead to measurably different behavior of the coupled system.
	Our numerical simulations therefore take into account the multi-level nature of the transmon and cavity, using experimentally determined parameters.
	In the rotating wave approximation, the Hamiltonian is given by:
	\begin{align}
		\H =& \Pi_{\omega} + \omega_{r} a\hc a + \sum_{i} g_{i} \l( \sigma_{+}^{(i)} a + \sigma_{-}^{(i)} a\hc \r) \nn\\
			&+ \frac12 \l( \epsilon_{ro} \ee^{-\ii \omega_{ro} t} a\hc + \epsilon_{ro}\cc \ee^{\ii \omega_{ro} t} a \r) 
				+ \frac12 \l( \epsilon_{q} \ee^{-\ii \omega_{d} t} a\hc + \epsilon_{q}\cc \ee^{\ii \omega_{d} t} a \r) \,,
		\label{eq:MultiJC}
	\end{align}
	where $\Pi_{\omega} = \sum_{i} \omega_{i} \ket i \bra i$ describes the transmon qubit in its eigenbasis with eigenstates $\ket i$ and corresponding energies $\omega_{i}$. 
	The qubit multi-level ladder operators are defined as $\sigma_{+}^{(i)} = \ket{i+1}\bra i$ and $\sigma_{-}^{(i)} = {\sigma_{+}^{(i)}}\hc$.
	The cavity has frequency $\omega_{r}$ and is coupled with the coupling strengths $g_{i}$ to the qubit transitions via a Jaynes-Cummings-type interaction.
	Each of the $g_{i}$ can be determined from the dipole moments corresponding to different qubit transitions, which are obtained by diagonalizing the transmon qubit Hamiltonian in the charge basis \cite{Koch2007}.
	The cavity measurement drive is applied at frequency $\omega_{ro}$ and with amplitude $\epsilon_{ro}$.  In the experiment, driving of the qubit is achieved as a second order effect via off-resonant driving of the cavity at the desired qubit transition frequency; 
	therefore in our model the qubit is driven indirectly via the cavity, with drive frequency $\omega_{d}$ and drive strength $\epsilon_{q}$.
	In the dispersive limit, $\l( \omega_{r} - \omega_{10} \r) = \Delta \gg g$, this term can be written perturbatively as the directly driven model $\sim \sum_{i} \Omega_{R,i} \l( \sigma_{+}^{(i)} + \sigma_{-}^{(i)} \r)$, 
	with the Rabi frequencies for each qubit transition $\Omega_{R,i} \approx \epsilon_{q} g_{i} / \Delta$. 
	However, we use the exact model Eq.~\eqref{eq:MultiJC} for our simulations. 
	
	To calculate the time evolution of the system under the influence of the environment, we write the master equation for the density matrix as~\cite{Boissonneault2010}
	\begin{align}
		\dot \rho &= \mathcal{L} \rho = -\ii \comm{\H}{\rho} + \kappa \diss{ a} \rho + 2\gamma_{\varphi} \diss{\Pi_{\delta \Phi}}\rho + \sum_{i} \gamma_{i} \diss{\sigma_{-}^{(i)}}\rho \, ,
		\label{eq:MEDet}
	\end{align}
	with photon loss from the cavity at rate $\kappa$, qubit decay from state $\ket i$ at rate $\gamma_{i}$ 
	and pure dephasing acting on the transmon due to fluctuations in its energy levels at rate $\gamma_{\varphi}$.
	Here $\Pi_{\delta \Phi} = \sum_{i} \delta \Phi_{i} \ket i \bra i$ and $\delta \Phi_{i} = \frac{\partial (\omega_{i} - \omega_{0 })}{\partial \Phi} \times \left( \frac{\partial (\omega_{1} - \omega_{0 })}{\partial \Phi} \right)^{-1}$, 
	where $\Phi$ is the external magnetic flux applied to the qubit loop. Then $\diss{\Pi_{\delta \Phi}}\rho$ describes dephasing of the transmon due to slow fluctuations in the magnetic field through the qubit loop.
	As decay of superconducting qubits is mostly due to electromagnetic coupling to spurious environmental modes~\cite{Barends:2013, Mueller:2015}, the relevant matrix elements are proportional to the respective dipole elements of the state transitions. 
	We therefore define the relaxation rates of higher transmon levels as $\gamma_{i} = \l( g_{i} / g_{0} \r)^{2} \gamma_{1}$, where $\gamma_{1}$ is the relaxation rate of the first excited qubit level. 
	All rates used in the simulations were measured in independent experiments.
	
	To simulate the measurement process, we use a stochastic master equation technique~\cite{Gambetta2008, Wiseman:1993a, Wiseman:book}.
	For a homodyne measurement, we write the stochastic master equation as
	\begin{align}
		\dot \rho = \mathcal L \rho + \sqrt{\kappa \eta} \xi(t) \l( \op c \rho + \rho \op c\hc - \mean{\op c + \op c\hc} \rho \r) \,, 
	\end{align}
	where $\op c = \ee^{\ii \varphi} a$ is the measurement operator for our system with the local oscillator phase $\varphi$ (with respect to the cavity driving signal), 
	$\kappa$ is the measurement rate (due to leakage out of the cavity) and $\eta$ is the measurement efficiency.  Here $\xi(t)$ is a stochastic noise process defined by $\xi(t) = dW(t) / dt$, where $dW(t)$ is the Wiener increment 
	and $dt$ is the timestep between successive evaluations of the master equation. 
	$dW$ represents the measurement noise and has the properties 
	\begin{align}
		\mean{dW}_{E} = 0 \quad \quad \mean{dW^{2}}_{E} = dt\; ,
	\end{align}
	where the average is over different realizations of the noise process.
	We then use the simulated stochastic dynamics to numerically calculate the qubit populations $\mean{\ket i \bra i}_{t}$ 
	and determine state transition rates with the same algorithm used on the experimental data. 
	Here the average is defined as $\mean{\op o}_{t} = \text{Tr}\l\{ \rho(t) \op o \r\}$.
	We also calculate the cavity field amplitude $\mean{a}_{t}$, the photon number $\mean{ a\hc a }_{t}$, and the associated ac Stark shifts to calibrate the measurement strength.  
	
	The numerical effort involved in these simulations is quite large, as the strong measurement and qubit drive requires one to take into account up to five levels of the transmon qubit as well as up to 50 photon states to ensure proper convergence. 
	Also, the time steps used in the stochastic numerical simulations has to be chosen much smaller than any of the intrinsic timescales of the problem to ensure that no unphysical solutions will be reached. 
	Due to these restrictions, the number of simulated data traces we used for comparing numerical and experimental transition rates is much smaller than the number of experimental traces. 
	However, since in the numerics we have direct access to the qubit populations, instead of having to infer them from the measurement record, convergence of the rate extraction algorithm is much faster for this data. 

\section{Multi-level driven Purcell effect \label{purcappend}}

	\begin{figure}[tbp]
	\begin{center}
	\includegraphics[width=3in]{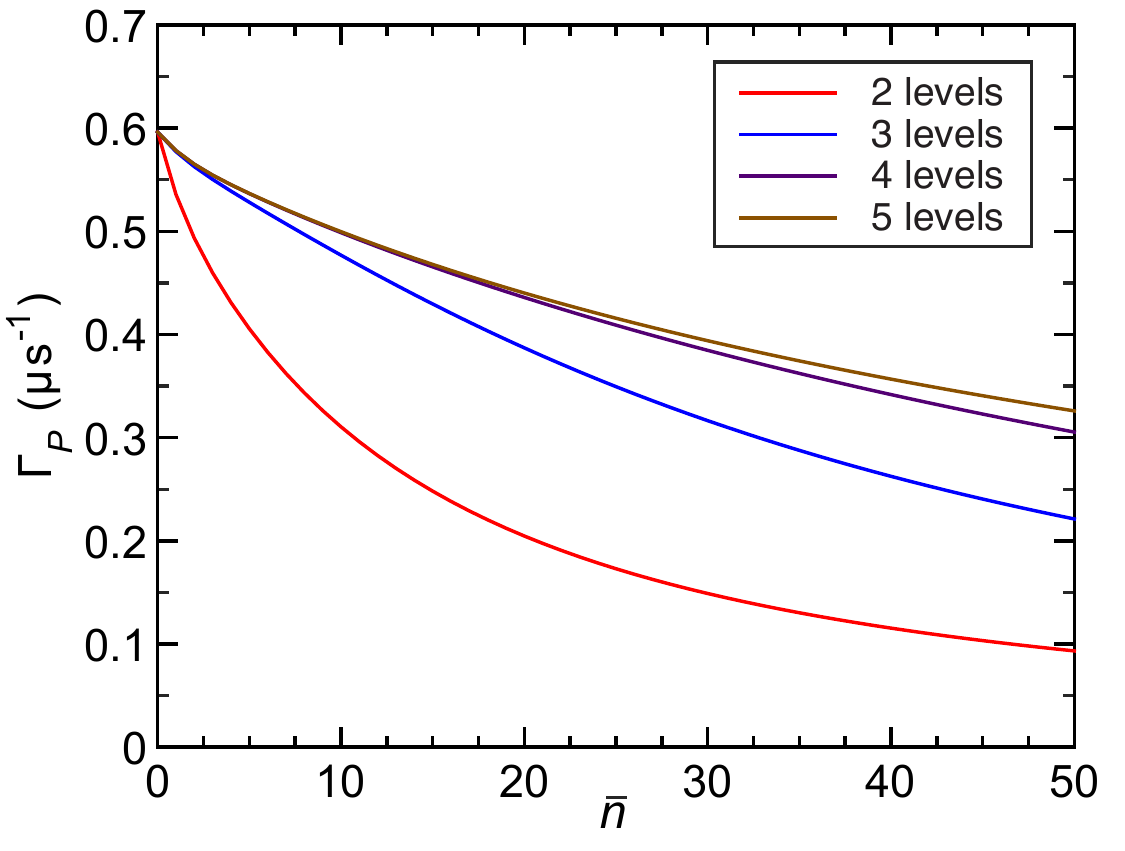}
	\end{center}
	\caption{Calculated purcell decay rate $\Gamma_P(\bar{n})$ for a coherent state with mean photon number $\bar{n}$, from Eq. \eqref{purccoh}.  The different curves are calculated by numerical diagonalization of the Hamiltonian \eqref{purcham} with between two and five qubit levels considered. Plots are made using the same qubit and cavity parameters used in the numerical simulations.}
	\label{fig:GammaPurcell}
	\end{figure}

	The model Hamiltonian for the Purcell decay rate of a multilevel qubit in a strongly driven cavity is:
	\begin{align}
		\label{purcham}
		\H = \omega_{r} a\hc a + \sum_{i} \omega_{i} \proj{i} + \sum_{i} g_{i} \l( \sigma_{i}^{+} a + \sigma_{i}^{-} a\hc \r) \,,
	\end{align}
	where $\sigma_{i}^{+} = \ket{i+1}\bra{i}$ and the sum goes over all relevant qubit levels.  Following Refs.~\onlinecite{Boissonneault2010, Sete2014}, we define the qubit's Purcell decay rate with the cavity in a Fock state with photon number $n$ as
	\begin{align}
		\Gamma(n) = \kappa \abss{ \overline{\bra{g,n}} a \overline{\ket{e,n}} } \,,
	\end{align}
	where the eigenstates $\overline{\ket{g,n}}$ and $\overline{\ket{e,n}}$ belong to the subspaces of the Jaynes-Cummings Hamiltonian with $n$ and $n+1$ excitations, respectively.  If the cavity is not in a Fock state with definite photon number $n$, the effective Purcell rate can be found by averaging over the photon number distribution of the cavity state $P(n)$ as
	\begin{align}
		\label{purccoh}
		\Gamma_{P} = \sum_{n} P(n) \Gamma(n) / \sum_{n} P(n) \,.
	\end{align}
	For a coherent state with mean photon number $\bar{n}$, the photon number distribution is $P(n) = \ee^{-\bar n} \bar n^{n} / n!$.
	
	Since in the absence of dissipation the Jaynes-Cummings Hamiltonian preserves the number of excitations, 
	the task of calculating the Purcell rate reduces to diagonalizing the Hamiltonian for each $n$-excitation subspace.

	The general form of the Hamiltonian in the $n$-excitation subspace can be written as
	\begin{align}\label{todiag}
		\H^{(n)} = \l(
		\begin{array}{ccccc}
			0 & \sqrt{n} g_{0} & 0 & 0 & \ldots \\
			\sqrt{n} g_{0} & \delta\omega_{1} & \sqrt{n-1} g_{1} & 0 \\
			0 & \sqrt{n-1} g_{1} & \delta\omega_{2} & \sqrt{n-2} g_{2} \\
			0 & 0 & \sqrt{n-2} g_{2} & \delta\omega_{3} &  \\
			\ldots & & & & \ldots
		\end{array} \r) \,,
	\end{align}
	with the $\delta\omega_{k} = \omega_{k} - k \omega_{r}$ giving the detuning between the $k^\mathrm{th}$ qubit level and a $k$-photon Fock state.
	We can write the eigenstates of the $n$-excitation subspace in the basis of qubit-cavity product states as:
	\begin{align}
		\overline{\ket{g,n}} &= b_{g,n} \ket{g,n} + b_{e,n} \ket{e,n-1} + b_{f,n} \ket{f,n-2} + \ldots \,, \nn\\
		\overline{\ket{e,n-1}} &= c_{g,n} \ket{g,n} + c_{e,n} \ket{e,n-1} + c_{f,n} \ket{f,n-2} + \ldots \,,
	\end{align}
	and so on for all other states involving higher qubit levels.  Here $\ket{g}$, $\ket{e}$, and $\ket{f}$ denote the first three qubit eigenstates, 
	and the $b_{i,n}$ and $c_{i,n}$ are complex numbers calculated by diagonalizing Eq.~\eqref{todiag}.
	Together with the action of the photon annihilation operator on cavity Fock states, $a \ket n = \sqrt{n} \ket{n-1}$, and noting that $a$ will lower the overall excitation number by one, we can then identify the relevant matrix elements of $a$ as
	\begin{align}
		\overline{\bra{g,n}} a \overline{\ket{e,n}} &= \sqrt{n+1}\: b_{g,n} c_{g,n+1} + \sqrt{n} \: b_{e,n} c_{e,n+1} + \sqrt{n-1} \: b_{f,n} c_{f,n+1} + \ldots \,,
	\end{align}
	where we obtain the coefficients numerically, using the experimental parameters for our transmon qubit.  
	Fig.~\ref{fig:GammaPurcell} shows a plot of the calculated Purcell decay rates for our experimental parameters with a coherent state in the cavity, accounting for various numbers of qubit levels.  The Purcell decay is suppressed by the presence of photons in the cavity, although the effect is less pronounced when higher qubit levels are accounted for.  The calculated rates for more than five qubit levels (not plotted) are very similar to those for five qubit levels.

\section{\label{mleappend}Maximum likelihood expressions for rate estimation}

Our qubit is a two-state system where state transitions obey Poissonian statistics, but the finite bandwidth of the measurement apparatus\textemdash which results from the inherent bandwidth $\kappa$ of circuit QED measurement, the bandwidth of the superconducting parametric amplifier and other elements in the measurement chain, and the Gaussian filtering employed to increase SNR in post-processing\textemdash tends to skew the observed dwell times in each state toward longer times and reduces the number of events seen with dwell times smaller than the measurement bandwidth.  A mathematical model to compensate for these effects has been presented by Naaman and Aumentado \cite{Naaman2006}.

\begin{figure}[t]
\begin{center}
\includegraphics[width=3in]{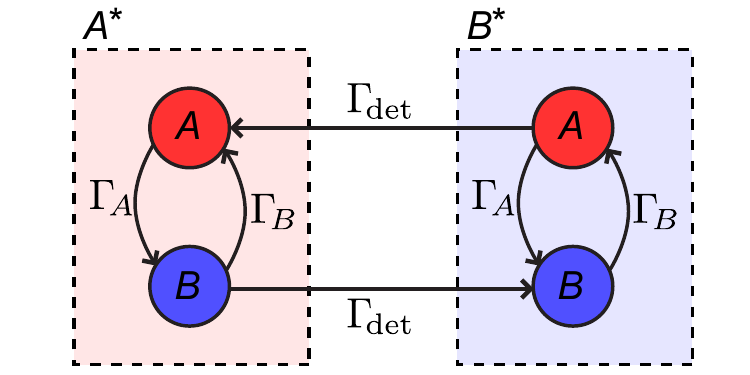}
\end{center}
\caption{\label{pdfmodel} State diagram for finite bandwidth detection.}  This figure shows a system state diagram for finite bandwidth detection.  The qubit is in either state $A$ or $B$ (circles), while the readout is in state $A^*$ or $B^*$.  For each qubit/readout state, the likelihood of transitioning to a different qubit/readout state is determined by the rates $\Gamma_A$, $\Gamma_B$, and $\Gamma_\mathrm{det}$ as indicated.  Adapted from ref. \cite{Naaman2006}.
\end{figure}

Their model allows the states of the qubit (denoted $A$ and $B$) to be independent of the corresponding states of the readout (denoted $A^*$ and $B^*$).  A diagram of this model, adapted from Ref. \onlinecite{Naaman2006}, is shown in Figure \ref{pdfmodel}.  For each readout state, the qubit can be in either state $A$ or $B$.  Transitions occur between $A$ and $B$ with rates $\Gamma_A$ and $\Gamma_B$.  However, when the qubit and readout ``disagree'', i.e. states $(B,A^*)$ and $(A,B^*)$, the readout can also make a transition (to $(B,B^*)$ or $(A,A^*)$, respectively) with rate $\Gamma_\mathrm{det}$.  Assuming that the readout does not change states unless the qubit has changed states (no false positives), one can write the probability distribution $h(t)$ for the observed dwell times in $A^*$ as a function of the underlying rates $\Gamma_A$, $\Gamma_B$, and $\Gamma_\mathrm{det}$ \cite{Naaman2006}:  

\beq
\label{joepdf}
h(t \,;\, \Gamma_A, \Gamma_B, \Gamma_\mathrm{det})=\frac{2}{\theta_A}\Gamma_A\Gamma_\mathrm{det} e^{-\lambda t/2}\sinh\left(\frac{\theta_A t}{2}\right),
\eeq
where $\lambda=\Gamma_A+\Gamma_B+\Gamma_\mathrm{det}$ and $\theta_A=\sqrt{\lambda^2-4\Gamma_A\Gamma_\mathrm{det}}$.  The probability distribution for dwell times in state $B^*$ is identical but with $\Gamma_A$ and $\Gamma_B$ interchanged (defining $\theta_B=\sqrt{\lambda^2-4\Gamma_B\Gamma_\mathrm{det}}$), as can be seen from the symmetry of the model.  We can calculate the complementary cumulative distribution, often called the survival function, by integrating the expression in \eqref{joepdf}:

\beq
\label{joesurv}
s(t \,;\, \Gamma_A, \Gamma_B, \Gamma_\mathrm{det})=\int_t^\infty h(t')dt'=\frac{e^{-\lambda t/2}}{\theta_A}\left[\lambda\sinh\left(\frac{\theta_A t}{2}\right)+\theta_A\cosh\left(\frac{\theta_A t}{2}\right)\right]
\eeq

The survival function is useful for handling right-censored dwell times, where the observed dwell time represents a lower bound on the true dwell time in the state.  The survival function $s(\tau)$ gives the probability of a dwell time $t>\tau$, which is precisely the probability distribution function needed for right-censored events.  Given a data set of $n_{A^*}$ dwell times in state $A^*$ and $n_{B^*}$ dwell times in state $B^*$, denoted $\{t_i\}$ and $\{t_j\}$, respectively, as well as corresponding censoring variables $\{\delta_i\}$ and $\{\delta_j\}$, where $\delta=0(1)$ indicates an uncensored (right-censored) dwell time, we can write the likelihood function as:  

\beq
\begin{split}
\mathcal{L}(\Gamma_{{A}},\Gamma_{{B}},\Gamma_{\mathrm{det}})=\prod_{i=1}^{n_{A^*}}h&(t_{i}\,;\,\Gamma_{{A}},\Gamma_{{B}},\Gamma_{\mathrm{det}})^{1-\delta_{i}}s(t_{i}\,;\,\Gamma_{{A}},\Gamma_{{B}},\Gamma_{\mathrm{det}})^{\delta_{i}}  \times \prod_{j=1}^{n_{B^*}}h(t_{j}\,;\,\Gamma_{{B}},\Gamma_{{A}},\Gamma_{\mathrm{det}})^{1-\delta_{j}}s(t_{j}\,;\,\Gamma_{{B}},\Gamma_{{A}},\Gamma_{\mathrm{det}})^{\delta_{j}} 
 \end{split}
\eeq

We then use nonlinear optimization methods to maximize the value of $\mathcal{L}$ by varying the parameters $\Gamma_A$, $\Gamma_B$, and $\Gamma_\mathrm{det}$.  Because of the limitations of floating point arithmetic, we actually perform the maximization on the log-likelihood function $L=\mathrm{ln}(\mathcal{L})$, which has the form:  

\beq
\begin{split}L(\Gamma_{{A}},\Gamma_{{B}},\Gamma_{\mathrm{det}})=\sum_{i=1}^{n_{A^*}}\Big[&(1-\delta_{i})\ln[h(t_{i}\,;\,\Gamma_{{A}},\Gamma_{{B}},\Gamma_{\mathrm{det}})]+\delta_{i}\ln[s(t_{i}\,;\,\Gamma_{{A}},\Gamma_{{B}},\Gamma_{\mathrm{det}})]\Big] \\
 & +\sum_{j=1}^{n_{B^*}}\Big[(1-\delta_{j})\ln[h(t_{j}\,;\,\Gamma_{{B}},\Gamma_{{A}},\Gamma_{\mathrm{det}})]+\delta_{j}\ln[s(t_{j}\,;\,\Gamma_{{B}},\Gamma_{{A}},\Gamma_{\mathrm{det}})]\Big]
\end{split}
\eeq

The functional forms of $\mathrm{ln}[h(t)]$ and $\mathrm{ln}[s(t)]$ for dwell times in state $A^*$ are:

\beq
\ln[h(t \,;\, \Gamma_A, \Gamma_B, \Gamma_\mathrm{det})]=\ln\left(\frac{2\Gamma_{{A}}\Gamma_{{B}}}{\theta_{{A}}}\right)-\frac{\lambda t}{2}+\ln\left(\sinh\left(\frac{\theta_{{A}}t}{2}\right)\right)
\eeq
\beq
\ln[s(t \,;\, \Gamma_A, \Gamma_B, \Gamma_\mathrm{det})]=-\ln(\theta_{{A}})-\frac{\lambda t}{2}+\ln\left[\lambda\sinh\left(\frac{\theta_{{A}}t}{2}\right)+\theta_{{A}}\cosh\left(\frac{\theta_{{A}}t}{2}\right)\right]
\eeq

The corresponding functions for dwell times in state $B^*$ can be found by interchanging $\Gamma_A$ with $\Gamma_B$, and $\theta_A$ with $\theta_B$.  For sufficiently large values of $\theta_A t$, these expressions may cause overflows in double-precision floating point arithmetic.  Therefore, for $\theta_A t > 40$, we make the following approximations, which introduce fractional errors of less than $10^{-18}$ for each approximated term:

\beq
\ln\left(\sinh\left(\frac{\theta_{{A}}t}{2}\right)\right)\approx\frac{\theta_{{A}}t}{2}-\ln 2
\eeq 

\beq
\ln\left[\lambda\sinh\left(\frac{\theta_{{A}}t}{2}\right)+\theta_{{A}}\cosh\left(\frac{\theta_{{A}}t}{2}\right)\right]\approx\frac{\theta_{{A}}t}{2}+\ln(\lambda+\theta_A)-\ln 2
\eeq

\end{document}